\documentclass[12pt,a4paper,final]{iopart}
\pdfoutput=1

\usepackage{amscd,amssymb}

\usepackage{epsfig}
\usepackage{amscd,amssymb}
\usepackage{graphicx,xspace}
\usepackage{pstricks}
\usepackage[left=3cm,top=3cm,right=3cm,bottom=3cm]{geometry}

\usepackage{cite}
\usepackage{bigfoot}
\usepackage{appendix}
\usepackage[breaklinks=true,colorlinks=true,linkcolor=black,urlcolor=black,citecolor=black,pdfencoding=auto]{hyperref}
\usepackage{amsfonts, amsmath, dsfont}
\usepackage{empheq}
\usepackage{bookmark}

\usepackage{color}
\definecolor{labelkey}{cmyk}{.4,.2,0,0}

\begin{document}

\newcommand \be  {\begin{equation}}
\newcommand \bea {\begin{eqnarray} \nonumber }
\newcommand \ee  {\end{equation}}
\newcommand \eea {\end{eqnarray}}

\title{Hessian spectrum at the global minimum of high-dimensional random landscapes}

\vskip 0.2cm

\author{Yan V Fyodorov}

\address{ King's College London, Department of Mathematics, London  WC2R 2LS, United Kingdom}

\author{Pierre Le Doussal}


\address{
Laboratoire de Physique Th\'eorique de l'Ecole Normale Sup\'erieure, PSL University
CNRS, Sorbonne Universit\'es, 24 rue Lhomond, 75231 Paris Cedex 05, France}

\begin{abstract}
Using the replica method we calculate the mean spectral density of the Hessian matrix at the global minimum of a random $N \gg 1$ dimensional isotropic, translationally invariant Gaussian random landscape confined by a parabolic potential with fixed curvature $\mu>0$.  Simple landscapes with generically a single minimum are typical for $\mu>\mu_{c}$, and we show that the Hessian at the global minimum is always {\it gapped}, with the low spectral edge being strictly positive.   When approaching from above the transitional point  $\mu= \mu_{c}$ separating simple  landscapes from ''glassy'' ones, with exponentially abundant minima, the spectral gap vanishes  as $(\mu-\mu_c)^2$. For $\mu<\mu_c$
the Hessian spectrum is qualitatively different for 'moderately complex' and 'genuinely complex' landscapes. The former are typical for short-range correlated random potentials and correspond to 1-step replica-symmetry breaking mechanism. Their Hessian spectra turn out to be again gapped, with the gap vanishing on approaching $\mu_c$ from below  with a larger critical exponent, as $(\mu_c-\mu)^4$. At the same time in the '' most complex'' landscapes with long-ranged power-law correlations the replica symmetry is completely broken. We show that in that case
  the  Hessian  remains gapless for all values of $\mu<\mu_c$, indicating the presence of 'marginally stable' spatial
 directions. Finally, the potentials with {\it logarithmic} correlations share both 1RSB nature and gapless spectrum.
  The spectral density of the Hessian always takes the semi-circular form, up to a shift and an amplitude that we explicitly calculate.

\end{abstract}

\maketitle

\section{Introduction}

\subsection{Formulation of the problem}
Understanding statistical structure of stationary points (minima, maxima and saddles) of random landscapes and fields of various types and dimensions is a rich problem of intrinsic current interest in various areas of pure and applied mathematics~\cite{AzaWsc09,Fyo15,AufBenCer13,AufBen13,Nic14,SubZei15,CamWig15,WaiTou13,FyoKho16,ChSch}.
It also keeps attracting steady interest in the theoretical physics community, and this over more than fifty years~\cite{Lon60,HalLax66,WeiHal82,Fre95,AnnCavGiaPar03,Fyo04,Par05,BraDea07,FyoWil07,FyoNad12}, with recent applications to statistical physics \cite{AnnCavGiaPar03,Fyo04,Par05,FyoWil07,FyoNad12,FyoLeD14,FLRT,RBBC2018}, neural networks and complex dynamics \cite{WaiTou13,FyoKho16,Fyo16}, string theory~\cite{DouShiZel04,DouShiZel06} and cosmology~\cite{EasGutMas16,YamVil2018}.

One of the most rich, generic and well-studied models in this class is described by the random function
\begin{equation}\label{landscape}
{\cal H}({\bf u})=\frac{\mu}{2}\sum_{k=1}^{N} u_k^2+V(u_1,...,u_{N}),
\end{equation}
where the curvature $\mu>0$ of the non-random confining parabolic potential is used to control the number of
 stationary points and $V({\bf u})$ is a mean-zero Gaussian-distributed random potential of ${\bf u}\in \mathbb{R}^N$ characterized by a particular (rotationally and translational invariant) covariance structure:
\begin{equation}\label{cov}
\overline{ V(\mathbf{u}_1) V(\mathbf{u}_2) } = N\: B\left(\frac{1}{2 N} (\mathbf{u}_1-\mathbf{u}_2)^2\right),
\end{equation}
In Eq.(\ref{cov}) and henceforth the notation
$\overline{ \cdots}$ stands for the quantities
averaged over the random potential, and $B(u)$ is a function of
order unity belonging to the so-called class ${\cal D}_{\infty}$
described in detail, e.g., in the book by Yaglom \cite{Yaglom}.
The functions $B(q)\in {\cal D}_{\infty}$ are such that they
 represent covariances of an isotropic random potential
 for {\it any} spatial dimension $N\ge 1$ (see
 a more detailed discussion below).

The mean number of stationary points and the mean number of minima for the random function ${\cal H}({\bf u})$ (which will be referred to as a 'landscape') was investigated  in the limit of large dimension $N \gg 1$ in \cite{Fyo04} and \cite{FyoWil07,FyoNad12}, see also \cite{Fyo15,BraDea07,YamVil2018}. In particular, it was found that a sharp transition  occurs from a 'simple' landscape for $\mu>\mu_{c}=\sqrt{B''(0)}$ with typically only a single stationary point (the minimum)  to a complex ('glassy') landscapes for $\mu<\mu_c$ with exponentially many stationary points, whose nature and statistics further depends on the properties of the covariance of the Gaussian function, see discussion below. Later on transitions of such type were suggested to be called the 'topology trivialization' transitions, see \cite{FyoLeD14,FLRT,Fyo16,RBBC2018} and references therein.   The mean number of stationary points
was also studied recently for the case $N=1$ \cite{FLRT} for the extended model of an elastic manifold of internal dimension $d$ embedded in dimension $N$,
where ${\bf u}$ is generalized to ${\bf u}(x)$, with $x \in \mathbb{R}^d$. In that context
the model (\ref{landscape}) is also known as the $d=0$ dimensional toy model
of a particle in a random landscape.  Note that
it also arises in the study of the Burgers equation with a random initial conditions in dimension $N$, which
exhibits interesting transitions and regimes, see e.g. for $N=1$ \cite{YanPLDBurgers}
and for large $N$ \cite{PLDMarkusBurgers}.

The goal of the present paper is to address for the model (\ref{landscape}) the density of eigenvalues of the $N\times N$ Hessian matrix
$K_{ij}({\bf u}_m)$, with ${\bf u}_m$ being the point of the {\it global minimum} of the landscape: ${\bf u}_m={\rm Argmin} \left[{\cal H}({\bf u})\right]$ and
\be \label{Hessian}
K_{ij}({\bf u})=\frac{\partial ^2}{\partial u_{i}\partial u_j}{\cal H}({\bf u})=\mu\delta_{ij}+\frac{\partial ^2}{\partial u_{i}\partial u_j} V({\bf u})
\ee
 being the landscape Hessian.

 This distinguishes our paper from earlier studies of this or closely related models where the statistics is sampled over all saddle-points or minima at a given value of the potential ${\cal H}({\bf u})=E=const$, see e.g.
\cite{RBBC2018,YamVil2018}.  In doing this we combine the random matrix theory with the methods of statistical mechanics of disordered systems, as described in detailed below. In this way we show that depending on the type of the covariance structure $B(u)$ as classified below, the spectrum of the (positive semidefinite) Hessian at the global minimum may be either 'gapped' (i.e. its lowest eigenvalue
is strictly positive in the limit $N\to \infty$), or 'gapless', when the lowest eigenvalue stays exactly zero. More precisely, the landscapes with generically a single minimum are typical for parameter $\mu$ exceeding a critical value $\mu_{c}$, and the Hessian at the global minimum of such simple landscape is always {\it gapped}.   When approaching the transition point  $\mu=\mu_{c}$ separating simple  landscapes from ''glassy'' ones with exponentially many stationary points, the gap in the Hessian spectrum vanishes quadratically. For $\mu<\mu_c$ the Hessian spectrum is qualitatively different for 'moderately complex' and 'genuinely complex' landscapes. The 'moderately complex' landscapes are typical for short-range correlated random potentials and, in the statistical mechanics context, are closely related to the 1-step replica-symmetry breaking (1RSB) mechanism. Their Hessian spectra at the global minimum turn out to be again gapped, with the gap  vanishing on approaching the transition point with a much higher exponent, see Fig. 1 below  for a plot of the lower edge of the Hessian for one example of a short-ranged potential.

 Let us now briefly outline a more quantitative description of our main results underlying the qualitative picture given above. We find that the mean (i.e. disorder-averaged) eigenvalue density $\rho(\lambda)$ for the Hessian at the global minimum is always given in the limit $N\to \infty$ by the semicircular law
\be\label{semisircleHessian}
 \rho(\lambda) = 8 \frac{ \sqrt{(\lambda_+-\lambda) (\lambda-\lambda_-)} }{\pi( \lambda_+-\lambda_-)^2} \quad , \quad
\int_{\lambda_-}^{\lambda_+} \rho(\lambda) d\lambda = 1
\ee
 the lower and upper spectral edges given by
\be\label{semicircleHessian1}
\lambda_{-} = \mu_{eff}- 2 \sqrt{B''(0)}, \quad \lambda_{+} = \mu_{eff}+ 2 \sqrt{B''(0)}
\ee
and $\mu_{eff}$ very essentially depends on the type of the covariance in Eq.(\ref{cov}).
For Short-Range correlated (SRC) potentials (represented, most importantly, by the family   $B(q)=C/(q+\epsilon)^{\gamma-1}$, $\gamma>1,\, C>0, \epsilon>0$)  the parameter $\mu_{eff}$ can be represented as
\be\label{mueffSRC}
\mu^{(SRC)}_{eff}=\mu+\frac{B''(0)}{\mu}+v\left(B'({\cal Q})-B'(0)-{\cal Q}B''(0)\right)
\ee
where the parameters $v>0,{\cal Q}\ge 0$ satisfy the system of equations
 \be\label{RSBQv1}
 \frac{\mu^2\,{\cal Q}}{1-\mu\,v\,{\cal Q}}=B'({\cal Q})-B'(0)
 \ee
 \be\label{RSBQv2}
 \frac{1}{v}\ln{\left(1-\mu\,v\,{\cal Q}\right)}=-\mu {\cal Q}+v\left[B({\cal Q})-B(0)-{\cal Q}B'({\cal Q})\right]
 \ee
which together with Eq.(\ref{mueffSRC}) can be shown to imply the following representation for the lower spectral edge
\be\label{lowedge1RSBintro}
  \lambda^{(SRC)}_{-}
=\left(\sqrt{\frac{\mu}{1- \mu v {\cal Q}}}-\sqrt{ \frac{1-\mu v {\cal Q}}{\mu}\, B''(0)}\right)^2 \ge 0
\ee
Note that the system of equations (\ref{RSBQv1})-(\ref{RSBQv2}) is obviously always satisfied by ${\cal Q}=0$ which defines the so called replica-symmetric solution. The solution is known to be correct (or 'stable') precisely in the phase with
a single minimum: $\mu>\mu_c=\sqrt{B''(0)}$. In that case $ \lambda^{(SRC)}_{-}=\frac{(\mu-\mu_c)^2}{\mu}$ and tends to zero quadratically on approaching $\mu_c$ from above. Below the transition for $\mu<\mu_c$ the system (\ref{RSBQv1})-(\ref{RSBQv2})
allows a solution with $Q>0$ which defines the phase with 1-step broken replica symmetry. For
 a general SRC covariance $B(q)$ the system can be solved only numerically, see an example in Fig.1.
   Close to the critical value however, when $\delta=1-\frac{\mu}{\mu_c}\ll 1$
 one can develop the perturbation theory in $0<\delta\ll 1$. The calculation presented in this paper shows that the first nonvanishing contribution to $ \lambda^{(SRC)}_{-}$ is order $O(\delta^4)$, and is given by
\be\label{lowedge1RSBperturb1}
  \lambda^{(SRC)}_{-}= \frac{\mu_c}{36[B'''(0)]^4}\left[\frac{3}{2}[B'''(0)]^2-B''(0)B^{''''}(0)\right]^2\delta^4+O(\delta^5)
 \ee

\begin{figure}
\centering
\includegraphics[scale=1.]{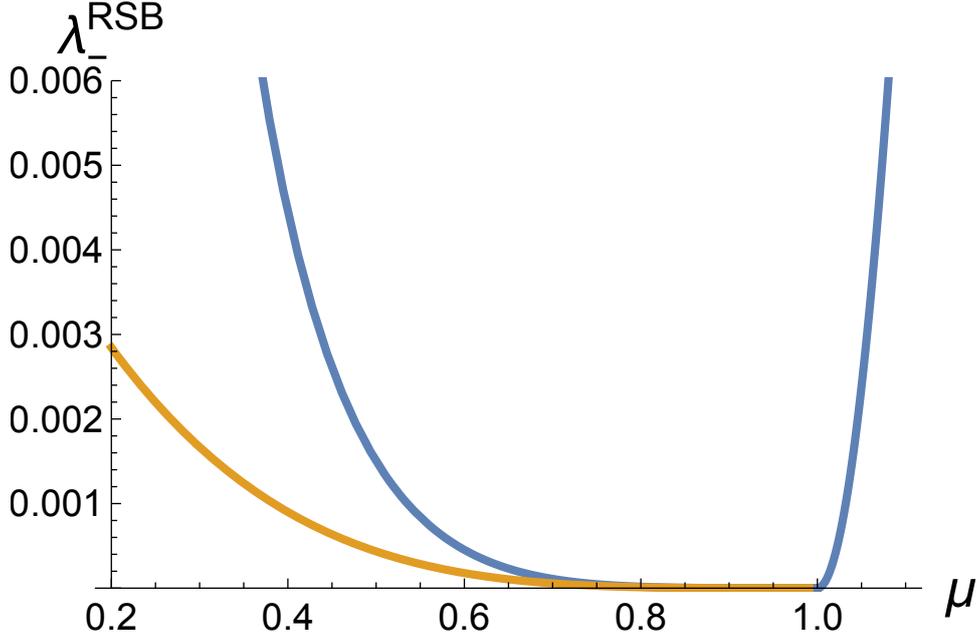}
\caption{ The lower edge of the Hessian spectrum $\lambda^{\rm (SRC)}_{-}$, associated to the 1 step replica symmetry breaking solution discussed in the text, plotted as a function of $\mu$ for the exponential SRC potential, with correlator $B(q)=e^{-q}$ (blue line for $\mu<1$). It vanishes
at the transition point for $\mu=\mu_c=1$ as $\sim (\mu_c-\mu)^4$ (orange line: the asymptote given by formula
\eqref{lowedge1RSBperturb1}). Also plotted (blue line for $\mu>1$) is the lower edge $\lambda^{\rm (SRC)}_{-}
= (\mu-1)^2/\mu$ for the replica symmetric solution above the transition point. The small $\mu$ behavior of
$\lambda^{\rm (1RSB)}_{-}$ (not shown here) is
discussed in the \ref{app2}.}
\label{fig1}
\end{figure}

The coefficient in front of the leading term is generically strictly positive unless $\frac{3}{2}[B'''(0)]^2-B''(0)B^{''''}(0)=0$. A more detailed study shows that the latter condition can be satisfied only for the logarithmically-correlated case with $B(0) - B(q) = C\ln{(1 + \frac{q}{\epsilon})}\,; C,\epsilon>0$. Moreover, we will find that such vanishing is not a coincidence: for the logarithmic case the spectral gap $ \lambda_{-}$ is {\it identically zero} everywhere in the phase with broken replica symmetry, $\mu<\mu_c$, so that the associated spectrum is {\it gapless}. Moreover, the property turns out to be shared by all the ' genuinely complex' landscapes generated by the so-called Long-Range Correlated
(LRC) potentials, most generically by those with the covariance $B(q)-B(0)= - C \left( (q+\epsilon)^{1-\gamma}- \epsilon^{1-\gamma} \right)$, with $0<\gamma<1$ and  $C,\epsilon>0$. In the latter case however the parameter $\mu_{eff}$ is not given by Eq.(\ref{mueffSRC}) but by a more involved expression
obtained in the framework of the Full Replica Symmetry Breaking scheme (FRSB) corresponding to a hierarchical picture of 'valleys within valleys within valleys'  discovered in the context of spin glass models \cite{MPV}. We show that in that case  the  Hessian at the global minimum  remains gapless for all values of $\mu<\mu_c$.

 Note that the properties of the Hessian at the global minimum determine the spectrum of the fluctuating modes in equilibrium, and is thus of fundamental importance. Clearly, the 'gapless' spectrum reflects an existence of very 'flat' directions in configuration space along which moving away from the global minimum incurs very little 'cost'. In the spin glass literature such flatness is known as a 'marginal stability' and is considered to be  one of the fundamental properties of the systems with the hierarchical structure of the energy landscapes, see  \cite{MPV} for early insights and \cite{MLC2006} for a detailed more recent discussion of this and related questions.
   In particular,  recently the 'marginal stability' has been shown to play the defining role in the spectrum of low-frequency modes of glasses \cite{FPUZ}. In the latter work the spectrum of the soft modes was calculated in a mean-field model of the jamming transition, the so-called 'soft spherical perceptron'. The Hessian matrix in that model has the shape of a (uniformly shifted) Wishart matrix, whose spectrum is given by the
(shifted) Marchenko-Pastur  law. The model has two phases: ' RS simple' and 'FRSB complex' and the Marchenko-Pastur spectrum  in that model was demonstrated to undergo a transition from gapped to gapless, similar to what we find here for Gaussian landscapes. Interestingly,
in our model the Hessian spectrum is given by a shifted Wigner semicircle, rather than the Marchenko-Pastur law, with shifts
and widths of the semicircle conspiring in an interesting way to produce the intricate picture just described.  We also note that the excitation spectrum was shown to be gapless in a {\it quantum} spin glass model at zero temperature \cite{AndMuel}
(see also e.g. \cite{TGPLDQuantum,LCTGPLDWigner,SchehrSpecificHeat,ChalkerGurarieBosons}).

Before describing our approach it is worth briefly mentioning an alternative model to Eq.(\ref{landscape}) which is popular in studies of high-dimensional Gaussian landscapes. This is the so-called {\it spherical $p-$spin} model \cite{CS}, which attracted a renewed attention recently \cite{AufBenCer13,AufBen13,SubZei15,FyoLeD14,Fyo16,RBBC2018}. In particular, early works \cite{KurLal996} and \cite{Cavagna1998} discussed Hessians for a variant of such model displaying transition from 'simple' to 'complex' landscapes of 1RSB type. Rather than concentrating on the global minimum, the works addressed the properties of the Hessians sampled over all stationary points at a given depth (called in \cite{KurLal996} the 'energy density') $e$ of the potential. In that context the  authors found a threshold density $e_{thres}$ such that Hessians for saddle points sampled at densities $e<e_{thres}$ are all positively gapped, whereas the gap tends to zero when approaching  $e=e_{thres}$ and becoming {\it negative} for $e>e_{thres}$. This picture suggested that all stationary points are minima for $e<e_{thres}$ but become saddle-points with unstable directions above the threshold. This conclusion was rigorously confirmed in \cite{AufBenCer13,AufBen13} and in the most recent studies \cite{SubZei15,RBBC2018}.

\subsection{Statistical Mechanics Approach to the global minimum }

In a useful interpretation, the model (\ref{landscape}) represents an energy landscape of a single classical particle in high-dimensional random potential, and one is interested in building the associated statistical mechanics model. This can be done by introducing the canonical partition function ${\cal Z}_{\beta}$
associated with the model at temperature $T$, with $\beta=1/T>0$ being the inverse
temperature, as
\be\label{partfun}
 {\cal Z}_{\beta}=\int_{\mathbb{R}^N} e^{- \beta {\cal H}({\bf u})} d{\bf u}, \quad  d{\bf u}=\prod_{i=1}^N du_i, \quad , \quad \beta =1/T  \,,
\ee
and considering the Boltzmann-Gibbs weights
$\pi_{\beta}({\bf u})={\cal Z}_{\beta}^{-1} e^{-\beta {\cal H}({\bf u})}$ associated with any configuration
${\bf u}$. This interpretation allows to study properties of the landscape at the global minimum.
One defines the thermal averaged value of any function $g({\bf u})$
as $\langle g({\bf u}) \rangle_T :=\int g({\bf u}) \pi_{\beta}({\bf u})d{\bf u}$.
In the zero-temperature limit $\beta\to \infty$ the Boltzmann-Gibbs weights concentrate on the set of globally minimal values of the energy function {\cal H}({\bf u}), so that
 for any well-behaving function $\langle g({\bf u}) \rangle_T$ should tend to the value of $g({\bf u}_m)$ evaluated at the argument ${\bf u}_m={\rm Argmin}\left[{\cal H}({\bf u})\right]$.
 Although this fact is valid in every disorder realization, in practice one concentrates on finding the disorder-averaged values
 $\overline{\left\langle g({\bf u}) \right\rangle}_T$. In particular, in this paper we choose the function $g({\bf u})$ as the resolvent of the Hessian:
 \be\label{resolvent}
g({\bf u}):=G(\lambda;{\bf u})=\frac{1}{N}\Tr\frac{1}{\lambda-K({\bf u})}, \quad \lambda \in \mathbb{C}
 \ee
From this we proceed to calculating the mean spectral density of the Hessian
eigenvalues ``at a temperature $T$'', defined as
\be \label{rhoT}
\rho_T(\lambda)=\frac{1}{\pi} \lim_{{\rm Im}\,\lambda\to 0^-}\, {\rm Im} \,
\overline{\left\langle G(\lambda,{\bf u})\right\rangle_T}
 \ee
 our final aim being to obtain the mean spectral density of Hessian eigenvalues at the
 absolute minimum by setting temperature to zero:
 \be
\rho(\lambda)= \lim_{T \to 0} \rho_T(\lambda) = \frac{1}{\pi} \lim_{{\rm Im}\,\lambda\to 0^-}\, {\rm Im} \,
\overline{G(\lambda,{\bf u}_m)}
 \ee

For our goals we find it most technically transparent to represent the resolvent
 using the formal {\it first} replica limit identity:
\be \label{start}
G(\lambda,{\bf u}) := \frac{1}{N}{\rm Tr} \frac{1}{\lambda- K({\bf u})}  = - 2 \lim_{m \to 0} \frac{1}{mN}
\partial_\lambda [ \det(\lambda - K({\bf u})) ]^{-\frac{m}{2}}
\ee
and use a representation of the inverse determinants in terms of $m$ replicated Gaussian integrals over $N-$component real-valued vectors $\phi_{\alpha}$, with $\alpha=1,\ldots,m$:
\be\label{1streplicanew}
G(\lambda,{\bf u})= - 2 \lim_{m \to 0} \frac{1}{mN} \partial_{\lambda}  \int_{\mathbb{R}^{Nm}}  \prod_{\alpha=1}^m d\phi_\alpha
e^{- \frac{i}{2} \lambda \sum_{\alpha=1}^m \phi_\alpha^2}
e^{ \frac{i}{2} \sum_{\alpha=1}^m\phi_\alpha \cdot K({\bf u}) \cdot \phi_\alpha  }
\ee
where we
set the factor $(\frac{i}{\pi})^{m/2} \to 1$ for $m=0$.
The problem therefore amounts to first calculating the disorder and thermal averaged value
\be\label{Gbetanew}
\!\!\!\!\!\!\!\!\!\!\!\!\!\!\! \overline{\left\langle G(\lambda, {\bf u})\right\rangle_T} = - 2 \lim_{m \to 0} \frac{1}{mN} \partial_{\lambda}  \int_{\mathbb{R}^{Nm}}  \prod_{\alpha=1}^m d\phi_\alpha
e^{- \frac{i}{2} \lambda \sum_{\alpha=1}^m \phi_\alpha^2}
\overline{ \langle e^{ \frac{i}{2} \sum_{\alpha=1}^m \phi_\alpha \cdot K({\bf u}) \cdot \phi_\alpha  }  \rangle_T }
\ee
where
\begin{equation}\label{thermavenew}
\langle e^{ \frac{i}{2} \sum_{\alpha=1}^m \phi_\alpha \cdot K({\bf u}) \cdot \phi_\alpha  }  \rangle_T =
{\cal Z}_{\beta}^{-1} \int_{\mathbb{R}^N}  \, d{\bf u} \,\, e^{ \frac{i}{2} \sum_{\alpha=1}^m \phi_\alpha \cdot K({\bf u}) \cdot \phi_\alpha -\beta {\cal H}({\bf u})}
\end{equation}
and then, by performing the zero-temperature limit, to capture the contribution from the global minimum only.

 The task of evaluating the expectation over the random variables contained in the random 'energy function' ${\cal H}({\bf u})$ in both the numerator and the denominator of the Gibbs measure in Eq.(\ref{thermavenew})
 is one of the central problems in the theory of disordered systems.
 Although the problem is meaningful and interesting for any $N$, we 
 will be
mainly concerned with the limit of large $N\gg 1$, where we will 
develop a systematic method of analysis.  For the model in question the associated methods are known, and we give the necessary background in the next two sections.

\subsection{Two classes of random potentials}

Before recalling the known results on the statistical mechanics of the model
for large $N$, let us first specify the types of random potential functions $V({\bf u})$, defined
for arbitrary $N$, that we study here.
An important characteristics is the behavior of the correlator $B(q)$ for large $q$, and we
will generally define $\gamma$ such that
\be \label{defgamma}
B(q) \sim q^{1-\gamma}
\ee as in \cite{MP1,MP2} (although more
general choices of functions are possible, as we now discuss).

According to Yaglom \cite{Yaglom}, there are two essentially different
types of Gaussian random potentials whose covariance function $B(q)$, defined in (\ref{cov}), belongs to the class ${\cal D}_{\infty}$.

\begin{itemize}

\item
The first type corresponds
to genuine isotropic random  potentials, and for those the covariances $B(q)$ are
characterized by a non-negative normalizable " spectral density
function" $\tilde{B}(k)\ge 0, \, k\ge 0$ in terms of which $B(q)$
is represented as ( see \cite{Yaglom} p.354)
\begin{equation}\label{shortranged}
B(q)=\int_0^{\infty}e^{-k^2 q} \tilde{B}(k) dk, \quad
B(0)=\int_0^{\infty}\tilde{B}(k)dk<\infty\,.
\end{equation}
In particular, $B(q)$ is decreasing and convex, i.e. satisfies
$B'(q)<0, B''(q)>0$ $\forall q\ge 0$, and in addition
$B'(q\to\infty)=0$. Here and below the number of dashes indicates
the order of derivatives taken.
In the physics literature such a class of random potentials is often called "short range correlated"
(SRC). The most important example in this class is the power-law correlated potential with $B(q)=C/(q+\epsilon)^{\gamma-1}$, $\gamma>1,\, C>0$,  which for $\epsilon>0$ corresponds to smooth potentials.
There are other interesting families listed in \cite{Yaglom}, the majority of which however produces very rough realizations of potentials. If however one insists (as is needed for our goals here, see below)
on the finiteness of at least two derivatives of $B(q)$ at $q=0$, then deserves mentioning also the case $B(q)=Ce^{-aq}$, $C>0,$ (which can be considered as a limiting case $\gamma\to \infty$ of the power-law correlated potentials), as well as the so-called Matern family
$B(q)=C(a q)^{\nu/2}K_{\nu}\left(\sqrt{aq}\right)$ with $ C>0,\, a>0$ and 'roughness parameter'
$\nu$, where $K_{\nu}(x)$ stands for the
modified Bessel (a.k.a. Macdonald)  function and the condition $\nu>k$ ensures that the process is at least
$k$ times differentiable.

\item

The second type of covariances occurs in the situation when the normalization integral
$\int_0^{\infty}\tilde{B}(k)dk$ diverges. It corresponds to so-called "long range correlated" (LRC)
random potentials with isotropic {\it increments},
also known as
 {\it locally} isotropic random fields, see e.g.
\cite{Yaglom}, p.438. The spectral density function $\tilde{B}(k)$ is again non-negative, and
 must now satisfy the condition $\int_0^{\infty}\frac{k^2}{k^2+1}\tilde{B}(k)dk<\infty$.  Such conditions allow one to prove that in any dimension $N\ge 1$ there exists a random potential
whose {\it structure function}
$\frac{1}{2} \overline{(V(0)-V({\bf u}))^2} =B(0)-B(q)$, $q=u^2/(2N)$, is well defined and given by
\begin{equation}\label{longranged}
B(0)-B(q)=\int_0^{\infty} dk (1-e^{-k^2q})\tilde{B}(k)+Au\,,
\quad A\ge 0\,.
\end{equation}
In what follows we will impose an additional requirement $B'(q\to
\infty)=0$, which ensures $A=0$. The most
widely-known example of the locally isotropic LRC random potential is the
so-called {\it self-similar} random potential, see \cite{Yaglom} p.
441, characterized by the spectral density
 $\tilde{B}(k>0)=k^{2\gamma-3},\, 0<\gamma<1$. The corresponding
covariance behaves as
\begin{equation} \label{beh}
B(q)- B(0) = C_{\gamma}q^{1-\gamma}\,.
\end{equation}
 which is the well-known example of the high-dimensional
analogue of the fractional Brownian motion (and reduces to the
Brownian motion for $N=1$ and $\gamma=1/2$). As such it is not smooth, but
can be made smooth by introducing a proper cutoff for $\tilde B(k)$ at large $\tilde k$,
without changing the behavior (\ref{beh}) at large $q$ (see below).
\end{itemize}
 It is important to note that for the Hessian entries $K_{ij}=\mu\delta_{ij}+\frac{\partial ^2}{\partial u_{i}\partial u_j} V({\bf u})$ to be well defined, the potential should be twice differentiable, hence
$B''(0)$ should exist. This is what we will assume here, for both classes of potentials. It corresponds to
random potentials usually considered in the theory of pinning, for which
the existence of a so-called Larkin regime, and Larkin scale, is related to $B''(0)$ being finite
(see \cite{MP1,MP2,TGPLDBragg1,TGPLDBragg1a,TGPLDBragg2,LDMW} and references therein).
In fact $\mu_c$ defined below is related to $B''(0)$, and defines the Larkin mass of the
problem. On the other hand, what matters for the physical applications (and defines the nature of
the low-temperature phase in the present model) is the {\it large-distance} behaviour of the random
potential, controlled by $B(q\gg 1)$. Hence we will always assume some small-scale cutoff parameter
$\epsilon>0$, see examples below, to remain in that class of sufficiently smooth potentials.
This is easily realizable in the class ${\cal D}_\infty$ by choosing $\tilde B(k)$  fast enough decaying at large $k$ with their
four lowest moments defined.

\subsection{Statistical Mechanics of disordered high-dimensional landscapes}
\label{sec:high}

Let us now discuss the statistical mechanics of the model in the
large $N$ limit. 
Such simple yet non-trivial models play the role of
a laboratory for developing the methods 
to deal with
problems of statistical mechanics where an interplay between
thermal fluctuations and 
quenched disorder is
essential. Paradigmatic examples
are
spin glasses \cite{MPV}, but similar effects occur
for polymers
in random environments, for
phase separating interfaces in random field models or for elastic
manifolds pinned by random impurities. In general, the presence of
quenched disorder leaves little choice but to employ the so-called
replica trick, which is a powerful heuristic way of extracting the averaged
free energy of the system from the moments of the partition function.
Note that considerable progress has been achieved in the last decade in developing the rigorous
aspects of that theory \cite{Bovier,Panchbook}. For the present model, a
detailed rigorous study was done by Klimovsky \cite{Klim2012}, so one may hope that the main results of the present paper
can be eventually made rigorous as well.

Technically, the problem amounts to calculating the ensemble
average of the equilibrium free energy $F=-T\ln{{\cal Z}\,}$.  The
replica trick allows one to represent integer moments $
\overline{{\cal Z}^n}$ of the partition function in a form of some
multivariable non-Gaussian integrals, and one then faces the usual
problem of finding ways  to evaluate these integrals explicitly enough to
be able to perform the replica limit $n\to 0$. In systems of high dimension
this can be done using saddle point methods, which allow
to employ the intricate pattern of spontaneous replica symmetry breaking (1RSB or FRSB)
discovered by Parisi \cite{Parisi} in the framework of
the Sherrington-Kirkpatrick model of spin glass with
infinite-range interactions.

In this broad context, the model Eq (\ref{landscape}) has 
a long history. First, in order to have {\it bona
fide} thermodynamic properties, the partition function
${\cal Z}_\beta$ of the model should be well-defined for any realization of the random
potential. This is achieved by the confining non-random part $V_{con}({\bf
x})=\frac{\mu}{2}{\bf x}^2$ which prohibits escape of
the particle to infinity.
The curvature $\mu>0$ then plays, together with the temperature $T$, the role of the main control
parameter, and one of standard goals of the theory
is to investigate the phase diagram in the $(\mu,T)$ plane.
Studies of the phase diagram in the limit of large $N$, and of detailed physical properties,
were performed in \cite{MP1,MP2,Engel,FS2007} for the statics,
and in \cite{FM,longtime}
for the aging dynamics (which does present
strong relations with the statics). More general
classes of confining potentials were considered in \cite{FyoWil07}.
The generalized model of elastic manifolds, which reduces to Eq (\ref{landscape})
for $d=0$, exhibits quite similar features and was also
much studied \cite{MP1,LCJKPLD96}, especially in the context of pinned vortex lattices
\cite{TGPLDBragg1,TGPLDBragg1a,TGPLDBragg2} and in comparison to the functional RG theory of pinning
\cite{PLDKWLargeNDetails,LDMW}.

From all these studies it is known that the nature of the low-temperature phase of
the model crucially depends on the behaviour of the covariance function $B(q)$ for large arguments,
i.e. the exponent $\gamma$ in (\ref{defgamma}) (see below for a more precise
statement), leading to the distinction 'short-ranged' correlated (SRC) or 'long-ranged' (LRC).
It turns out that for the model Eq (\ref{landscape}) this distinction coincides with the
two classes discussed in the previous section (note that for the more general model
of a manifold, not studied here, there is a critical value such that LRC holds for $\gamma <\gamma_c(d)$,
with $\gamma_c(0)=1$). We will thus use this denomination everywhere.
In the statistical mechanics of the model the SRC potentials correspond to a 1 step replica
symmetry breaking (1RSB) low temperature phase, while the LRC potentials lead
to the full RSB pattern (FRSB). The precise criterion to classify covariance structures at large $N$, as SRC vs LRC, was studied in much
detail in \cite{FS2007} for the model Eq (\ref{landscape}) (see also \cite{longtime}),
and for the more general manifold problem in \cite{LDMW} (see also \cite{MP1}), two works to which we refer extensively below.

The distinction between the two cases has been put in a nice compact form
in \cite{FS2007} (see also \footnote{This classification is also obtained, more
generally for manifolds, by requesting a monotonous order parameter from
the full RSB solution, see e.g. Eqs 8.16 in \cite{PLDKWLargeNDetails}.}).
It employs the
notion of the so-called Schwarzian derivative
$\{B'(q),q\}=-A(q)/[B''(q)]^2$, where $A(q)$ is expressed in terms
of $B(q)$ as
\begin{equation}\label{Schwarziancriterion}
A(q)=\frac{3}{2}\left[B'''(q)\right]^2-B''(q)B''''(q)\,.
\end{equation}
In terms of $A(q)$ it was demonstrated that
\begin{itemize}
\item any potential whose covariance function satisfies the condition\\ $A(q)>0\,\, \forall q\ge 0$ must have
the 1RSB low temperature phase. It is easy to check that such a
situation includes, in particular, all standard families of the SR
potentials with covariances listed after Eq.(\ref{shortranged}).
It is natural to conjecture that this property is characteristic
for all SR isotropic potentials with covariances defined via
Eq.(\ref{shortranged}).

\item Any potential whose correlation function satisfies the
condition\\
$A(q)<0\,\,\forall q\ge 0$ must necessarily exhibit the FRSB low
temperature phase. This condition holds for the standard LR
correlation functions of the type Eq.(\ref{longranged}), i.e. for
$B(q)-B(0)= -g^2 \left( (q+\epsilon)^{1-\gamma}- \epsilon^{1-\gamma} \right)$, with $0<\gamma<1$, $\epsilon>0$,
obtained from the choice $\tilde B(k)=\frac{2 g^2}{\Gamma(\gamma-1)} k^{2 \gamma-3} e^{-k^2 \epsilon}$.

It is natural to conjecture that typical smooth
LR random potentials with independent increments should be of this
type.
\end{itemize}

Clearly, the above criterion naturally singles out as a special
marginal case random potentials satisfying $A(q)=0$. The only
function satisfying this condition globally, i.e for all $q\ge 0$,
and satisfying also the requirement $B'(q\to \infty)=0$ is indeed
given by a {\it logarithmic} correlation function, of the type studied in \cite{CLD}:
 \begin{equation}\label{logcorr}
B(0) - B(q) = g^2\ln{(1 + \frac{q}{\epsilon})}\,,
\end{equation}
where $g$ and $\epsilon>0$ are given constants. Let us stress that
the expression Eq.(\ref{logcorr}) is a legitimate covariance function
belonging to the ${\cal D}_{\infty}$ class of LR locally isotropic
fields, Eq.(\ref{longranged}). Indeed, it
can be considered as a limiting case $\gamma\to 1$ of the smooth LCR
potential discussed above and corresponds to the
spectral density of the form
$\tilde{B}(k)=\frac{2g^2}{k}e^{-\epsilon\, k^2}$, which satisfies the
required condition
$\int_0^{\infty}\frac{k^2}{k^2+1}\tilde{B}(k)dk<\infty$.

As shown in several works \cite{FS2007,TGPLDBragg1,TGPLDBragg1a,TGPLDBragg2,longtime} such logarithmically-correlated potentials induce thermodynamics with features of both the
SR-1RSB and LR-FRSB regimes, and are singled out in many respects.
In particular, one can show that this case is the only \footnote{One may also consider a more general class of {\it multiscale} log-correlated fields sharing this feature, see \cite{FB2008a}.} infinite-dimensional model where
the transition temperature remains finite when the confining parameter $\mu\to 0$.
In fact, such a case can be considered as an (infinitely) high-dimensional representative of an intensively studied class of log-correlated random processes and fields in {\it low dimensions} $N=1,2$. These processes, characterized by logarithmic singularity on the diagonal of its covariance kernel \cite{CLD}, appear with surprising regularity in many different contexts, such as the statistical mechanics of branching random walks and polymers on trees \cite{DS1988,CMW}, and associated extremal value problems
 \cite{CLD,FB2008,FLDR,DRZ,SubagZeitouni,OstrRev} of great relevance from random matrix theory and number theory \cite{FyoKeat2014,ABBRS}, to probabilistic description of two-dimensional gravity \cite{KRV-DOZZ,Remy}. In fact, many generic features of the low dimensional log-correlated
 potential are naturally expected to be shared by its high-dimensional counterpart, the latter deserves to be studied seriously in every aspect.

In the next section we perform the analysis of Eqs.(\ref{Gbetanew}) and (\ref{thermavenew}) by employing the {\it second} replica trick, adapting the considerations in works \cite{FS2007,LDMW} to the Hessian problem, with due modifications.

\section{Replica analysis of the Hessian problem}

\subsection{Main idea of the calculation}

In the framework of the replica trick we represent the
normalization factor ${\cal Z}^{-1}_{\beta}$ in Eq.(\ref{thermavenew})
formally as $1/{\cal Z}_{\beta}=\lim_{n\to 0}{\cal Z}_{\beta}^{n-1}$
and treat the parameter $n$ before the limit as a positive integer. After this is done, averaging the product of
$n$ integrals over the Gaussian potential $V({\bf u})$ is an easy task. Before embarking
on the detailed calculation let us sketch the main idea.
We will rewrite the disorder average of Eq.(\ref{thermavenew})
\be \label{form1new}
\overline{ \langle e^{ \frac{i}{2} \sum_{\alpha=1}^m \phi_\alpha \cdot K({\bf u}) \cdot \phi_\alpha  }  \rangle_T }
= \lim_{n \to 0}  \int \, \prod_{a=1}^n d{\bf u}_a \, e^{- L_{n,m}[\bf u, \phi] }
\ee
in terms of some replicated action $L_{n,m}[\bf u, \phi]$ that we will calculate. This allows to write (\ref{Gbetanew}) as
\begin{equation}\label{Gbeta2new}
 \overline{\left\langle G(\lambda, {\bf u})\right\rangle_T} =  \lim_{m \to 0, n \to 0}
 \int_{\mathbb{R}^{N(n+m)}}  \, \prod_{\alpha=1}^m d\phi_\alpha \prod_{a=1}^n d{\bf u}_a \,
 \frac{i \sum_{\alpha=1}^m \phi_\alpha^2}{m N} e^{- \frac{i}{2} \lambda \sum_{\alpha=1}^m \phi_\alpha^2}
e^{- L_{m,n}[\bf u, \phi] }
\end{equation}
where we have performed the derivative w.r.t. $\lambda$.
Note that formally setting $\frac{i \sum_{\alpha=1}^m \phi_\alpha^2}{m N} \to 1$ in (\ref{Gbeta2new})
is equivalent to setting $\frac{-2}{m N} \partial_\lambda{ =1}$ in the r.h.s of (\ref{start}),
which gives unity. Hence the integral in (\ref{Gbeta2new}) is normalized to unity
for $n=m=0$, and
we can rewrite (\ref{Gbeta2new}) as an expectation value
in the replicated theory
\be\label{Gl10}
\overline{\left\langle G(\lambda, {\bf u})\right\rangle_T} =  \lim_{m \to 0, n \to 0}
\langle \frac{i \sum_{\alpha=1}^m \phi_\alpha^2}{m N} \rangle_{n,m}
\ee
where $\langle {\cal O} \rangle_{n,m}$ denotes the expectation value of ${\cal O}$
in the theory with $n \times m$ replicated fields ${\bf u}_a,\phi_\alpha$
and the action $L_{n,m}[\bf u, \phi]$. Below we perform the
calculation of this expectation value at large $N$ using the saddle point method.

\subsection{Calculation}

We have
at the first step, from (\ref{thermavenew}) and (\ref{Hessian}):
\bea \label{thermave1new}
\fl && \overline{ \langle e^{ \frac{i}{2} \sum_{\alpha=1}^m \phi_\alpha \cdot K({\bf u}) \cdot \phi_\alpha  }  \rangle_T }
= \lim_{n \to 0}
e^{\frac{i \mu}{2} \sum_{\alpha=1}^m \phi_\alpha^2}\,
\int_{\mathbb{R}^{Nn}}  \,\prod_{a=1}^n d{\bf u}_a
\, e^{ - \frac{\beta\mu}{2}\sum_{a=1}^n {\bf u}_a^2} \\
\fl &&
~~~~~~~~~~~~~~~~~~~~~~~~~~~~~~~~~~~~~~~
\times e^{\frac{1}{2} \overline{ \left[-\beta \sum_{a=1}^n V({\bf u}_a)
+\frac{i}{2}\sum_{ij=1}^N \sum_{\alpha=1}^m \phi_\alpha^i \phi_\alpha^j
\frac{\partial^2}{\partial u_{1i}\partial u_{1j}}V({\bf u}_1)\right]^2}}
\eea
Now, by differentiating in Eq.(\ref{cov}) one find the following relations:
\begin{equation}\label{covcons1new}
 \left\langle V(\mathbf{u}) \frac{\partial^2}{\partial u_i\partial u_j}V({\bf u})\right\rangle =\delta_{ij} B'(0)
\end{equation}
\begin{equation}\label{covcons2new}
 \left\langle  \frac{\partial^2}{\partial u_i\partial u_j}V({\bf u}) \frac{\partial^2}{\partial u_k\partial u_l}V({\bf u})\right\rangle  =\frac{1}{N}B''(0)\left(\delta_{ij}\delta_{lk}+\delta_{ik}\delta_{jl}+\delta_{il}\delta_{jk} \right)
\end{equation}
and for $a\ne 1$:
\bea
\label{covcons3new}
\fl && \left\langle V(\mathbf{u}_a) \frac{\partial^2}{\partial u_{1i}\partial u_{1j}}V({\bf u}_1)\right\rangle =\delta_{ij} B'\left(\frac{\left({\bf u}_a-{\bf u}_1\right)^2}{2N}\right) \\
\fl && ~~~~~~~~~~~~~~~~~~~~~~~~~~~~~~~~~~~~~~~~~~~~~~~ +\frac{\left({\bf u}_{ai}-{\bf u}_{1i}\right)\left({\bf u}_{aj}-{\bf u}_{1j}\right)}{N}B''\left(\frac{\left({\bf u}_a-{\bf u}_1\right)^2}{2N}\right)
\eea
which, brings Eq.(\ref{thermave1new}) to the form
Eq.(\ref{form1new}) with
\begin{equation}\label{actionnew}
L_{n,m}[{\bf u}, \phi]={\cal L}_{m}[\phi]+{\cal L}_{n,m}[{\bf u}, \phi]
\end{equation} where the
${\bf u}$-independent part of the action is given by
\begin{equation}\label{thermave1Anew}
{\cal L}_{m}[\phi]=\frac{B''(0)}{8N}\left[ (\sum_{\alpha=1}^m \phi_\alpha^2)^2
+2 \sum_{\alpha,\beta=1}^m (\phi_\alpha \cdot \phi_\beta)^2 \right]
-\frac{i \mu}{2} \sum_\alpha \phi_\alpha^2
\end{equation}
whereas both
${\bf u}$- and $\phi-$ dependent part is
\begin{equation}\label{thermave1A2new}
\fl {\cal L}_{n,m}[{\bf u},\phi]=\frac{\beta\mu}{2}\sum_{a=1}^n {\bf u}_a^2
-N\frac{\beta^2}{2}\sum_{a,b=1}^n B\left(\frac{\left({\bf u}_a-{\bf u}_b\right)^2}{2 N}\right)
\end{equation}
\[
 +\frac{i\beta}{2} (\sum_\alpha \phi_\alpha^2)\,\sum_{a=1}^n B'\left(\frac{\left({\bf u}_a-{\bf u}_1\right)^2}{2 N}\right)
+\frac{i\beta}{2 N}\sum_{a=1}^nB''\left(\frac{({\bf u}_a-{\bf u}_1)^2}{2 N}\right)
\sum_{\alpha=1}^m
\left( \left({\bf u}_a-{\bf u}_1\right) \cdot \phi_\alpha \right)^2
\]

\medskip

Let us now examine the expression (\ref{Gbeta2new}) for
$\overline{\left\langle G(\lambda, {\bf u})\right\rangle_T}$
which has the form of an integral over $\mathbb{R}^{N(n+m)}$ for
the $n+m$ $N$-component vectors ${\bf u}_a$ and $\phi_\alpha$.
We see that,
given the expression (\ref{actionnew}),(\ref{thermave1Anew}),(\ref{thermave1A2new}) for the replicated action
$L_{n,m}[{\bf u}, \phi]$ the integrand in (\ref{Gbeta2new}) depends only on
the vectors ${\bf u}_a$ and $\phi_\alpha$ through the combinations
\be
Q_{ab}={\bf u}_a \cdot {\bf u}_b \quad , \quad
P_{\alpha \beta} = \phi_\alpha \cdot \phi_\beta \quad , \quad
R_{a \alpha}= {\bf u}_a \cdot \phi_\alpha
\ee
We can embody these scalar products into a $(n+m) \times (n+m)$ matrix $M$
of scalar products of combined vectors $({\bf u}_a,\phi_\alpha)$ as
\begin{equation}\label{M}
M =
 \left(\begin{array}{cc} Q   &  R \\
R^T & P
\end{array}\right)\ge 0 ,
\end{equation}

We can thus trade the integration variables ${\bf u}_a$ and $\phi_\alpha$ for
the matrix $M$ by using the invariant integration theorem, recalled in the
Appendix A, upon the change $n \to n+m$, ${\bf u}_a \to ({\bf u}_a,\phi_\alpha)$
and $Q \to M$ of the formula there. We also perform a final
rescaling $Q,R,P \to NQ,NR,NP$ and obtain
\be\label{Gl1new}
\fl \overline{\left\langle G(\lambda, {\bf u})\right\rangle_T} =  \lim_{m \to 0, n \to 0}
\, {\cal I}\left[ \frac{i \tr P}{m} \right]
\ee
where we denote the integral
\be \label{Gl1new2}
{\cal I}[ {\cal O} ]  = {\cal C}^{(o)}_{N,n+m} N^{\frac{n+m}{2} N}
 \int_{M \geq 0}  dM \, {\cal O}(M) \,
  e^{- N L(M) }(\det{M})^{- \frac{n+m+1}{2}}
\ee
of any observable ${\cal O}(M)$, which is now over the space of symmetric real positive semi-definite matrix $M$
with measure $dM = dQ dP dR$ where
\be
dQ=\prod_{1 \leq a \leq b \leq n} dQ_{ab} \quad , \quad
dP=\prod_{1 \leq \alpha \leq \beta \leq m} dP_{\alpha \beta}
\quad , \quad
dR=\prod_{a=1}^n \prod_{\alpha=1}^m dR_{a \alpha}
\ee
Note that $Q$ and $P$ are also symmetric real positive semi-definite matrices (we can consider them in fact positive definite as
the manifold of exactly zero eigenvalues forms the boundary of the integration domain and is therefore of lower dimension. Such boundaries can not contribute to the integral unless the integrand is manifestly divergent when approaching the boundary manifold.  One can check {\it aposteriori}
this is not the case in our problem). The off-diagonal block $R$ is a general real matrix such that $M$ remains positive definite. The latter condition imposing certain restrictions on the maximal singular value of $R$ (see about the singular values below) is however irrelevant for our consideration below.
The prefactor is given by (see the Appendix A)
\be
{\cal C}^{(o)}_{N,n+m} = \frac{\pi^{\frac{n+m}{2}\left(N-\frac{n+m-1}{2}\right)}}
{\prod_{k=0}^{n+m-1}\Gamma\left(\frac{N-k}{2}\right)}
\ee
We have also defined the action
\be
L[M]= L_m[P] + L_{n,m}[P,Q,R]
\ee
\be\label{actPnew}
L_m[P]=  \frac{B''(0)}{8}\left[\left(\tr\,P\right)^2+2\tr\left(P^2\right)\right]+\frac{i (\lambda-\mu)}{2} \tr P
\ee
\begin{equation}\label{actQnew}
L_{n,m}[P,Q,R]=-\frac{1}{2}\tr \log{M}+\frac{\beta\mu}{2}\tr Q-\frac{\beta^2}{2}
\sum_{a,b=1}^n B\left(\frac{Q_{aa}+Q_{bb}-2Q_{ab}}{2}\right)
\end{equation}
\[
+\frac{i\beta}{2} \tr P\,\sum_{a=1}^n B'\left(\frac{Q_{aa}+Q_{a1}-2Q_{11}}{2}\right)
\]
\be\label{lastpiece}
+ \frac{i\beta}{2}\sum_{a=1}^n B''\left(\frac{Q_{aa}+Q_{a1}-2Q_{11}}{2}\right)
\left( (R R^T)_{aa} + (R R^T)_{11} - 2 (R R^T)_{a1} \right)
\ee
The above expression is exact for any $N > n+m$.

 Note again that, by the same argument as below Eq. (\ref{Gbeta2new})
applied to (\ref{Gl1new}) shows that $\lim_{m \to 0, n \to 0}{\cal I}[ 1 ] = 1$. Hence we can write again
the mean resolvent as an expectation value
\be\label{Gl1new3}
\fl \overline{\left\langle G(\lambda, {\bf u})\right\rangle_T} =  \lim_{m \to 0, n \to 0} \,
\langle \frac{i \tr P}{m} \rangle_{L}
\ee
w.r.t. the action $L(M)$, where the expectation value of ${\cal O}$ is defined as
$\langle {\cal O} \rangle_{L}=
{\cal I}[ {\cal O} ]/{\cal I}[1 ]$.

Finally, note that the integrand in Eqs.(\ref{Gl1new}-\ref{Gl1new2}), including
the action $L(M)$, is invariant by the transformation
\be \label{invariance}
P \to {\cal O}^{T} P \, {\cal O} \quad , R \to R \, {\cal O}
\ee
for any orthogonal matrix ${\cal O}$ in $O(m)$ (a consequence of the original
invariance upon rotating the $m$ replica $\phi_\alpha$).

\subsection{Saddle point solution and the semi-circle density of Hessian eigenvalues}

The form of the integrand in Eqs.(\ref{Gl1new}) is suggestive of evaluating
the corresponding $P-$, $Q-$ and $R$ integrals by the saddle-point method for $N\gg 1$,
assuming, as usual, commutativity of the replica limit(s) with $N\to \infty$.
Hence we will now search for the stationary points, $M^{\rm sp}$, of $L(M)$. To leading
order as $N \to +\infty$ we will replace Eq.(\ref{Gl1new3})
by its value at the stationary points
\be\label{Gl1new3}
\fl \overline{\left\langle G(\lambda, {\bf u})\right\rangle_T} =  \lim_{m \to 0, n \to 0}
\frac{i \tr P^{\rm sp}}{m}
\ee
where $P^{\rm sp}$ is the $P$- component of the matrix $M^{\rm sp}$
 (note that because of symmetries such as \eqref{invariance}, there is a manifold of
stationary points, however the value of $\tr P^{\rm sp}$ is the same for all these stationary points).
Since Eq.(\ref{Gl1new3}) is an expectation value, we do not need to
calculate explicitly below all prefactors in the saddle point evaluation, and only
the value of $\tr P$ on the saddle point manifold matters to the leading
order at large $N$ (the prefactors would matter if one aimed to carry out a $1/N$ expansion).


Although $L(M)$ looks complicated, the first simplifying feature is that one can argue that the saddle-point solution corresponds to
the choice $R=0$.
Indeed, the saddle point equation, expressed along the $R$ components of $M$,
$\frac{\delta}{\delta R} L =0$,
is of the form, schematically
\be
(M^{-1})_{a \alpha} =  R
\ee
and since the inversion of Eq.(\ref{M}) leads to $(M^{-1})_{a \alpha} \sim R$ at small $R$, the choice $R=0$ always satisfies the saddle-point equation.
 To show that the solution $R=0$ is the relevant one, one needs to study the stability of the quadratic form by expanding to second order in $R$:
We have
\be
-\frac{1}{2}\tr \log{M} = -\frac{1}{2}\tr \log  \left(\begin{array}{cc} Q   &  0 \\
0 & P
\end{array}\right)
-\frac{1}{2}\tr \log\left( I +   \left(\begin{array}{cc} 0   &  R P^{-1} \\
R^T Q^{-1} & 0
\end{array} \right)\right)
\ee
\be \label{expand}
= -\frac{1}{2}\tr \log{Q} -\frac{1}{2}\tr \log{P} + \frac{1}{2} \tr R P^{-1} R^T Q^{-1} + O(R^4)
\ee
Hence this part of the action starts at the terms of second order in entries of $R$ around $R=0$, and the stability matrix for $M$ around the $R=0$ saddle point decouples  (i.e. there are no quadratic terms of the
kind $QR$ or $PR$). To check stability of $R=0$ solution one therefore only needs to check that the second-order terms  in  Eq.(\ref{expand}) can be represented as a positive definite quadratic form
 in the components of $R$ (note that the only other $R$-dependent term in $L(M)$, i.e. the last term of
Eq.(\ref{lastpiece}), being purely imaginary cannot modify the stability properties with respect to fluctuations in $R$).

In checking this consider first a special case $m=n$ such that $R$ is a square matrix.  Any such $R$ can be represented as $R={\cal O}_1 R_d {\cal O}_2^T$, where
where ${\cal O}_1\in O(n)$ and ${\cal O}_2\in O(n)$ are two different orthogonal  matrices, diagonalizing $RR^T$ and $R^TR$, correspondingly, and $R_d=diag(r_1, r_2,...,r_n)$, with $r_{a}\ge 0,\, a=1,\ldots,n$ being the so-called singular values of $R$. Defining ${\cal O}_1^TQ^{-1}{\cal O}_1=A>0$ and $B= {\cal O}_2^TP^{-1}{\cal O}_2 > 0$ we have $$\tr R^T Q^{-1} R P^{-1}=\tr R_d A B R_d = \sum_{ab}r_a r_b C_{ab}\,,$$ where $C_{ab}=A_{ab}B_{ab}$ (we used $B^T=B$).
Matrices $C$ of this form are known as the 'Hadamard product' of $A$ and $B$, and  according to  the 'Schur product theorem'
one has $C>0$ as long as $A> 0$ and $B>0$, showing that the associated quadratic form in singular values is positive definite.
As the result the value dominating the $R-$integration always corresponds to $R_d=0$, hence to the whole matrix $R=0$.
The same consideration can be straighforwardly extended to $n\ne m$ case, via representing the $m\times n$ matrix $R$  via a very similar
 singular value decomposition with  ${\cal O}_2\in O(m)$ and $R_d$ being $m\times n$ diagonal matrix which is zero at the saddle-point.  Thus, from now on we may assume $R$ to be integrated out in the quadratic approximation, i.e. we set $R=0$ in the saddle point and deal with only the diagonal blocks $Q$ and $P$ (as noted above we do not need, to the leading order at large $N$, to keep track of
 the pre-exponential factors produced by the $R$ integration).

It is now useful to observe that the form of the $P-$dependence in the integrand
Eqs.(\ref{Gl1new})-(\ref{Gl1new2}), invariant under Eq.(\ref{invariance}),
suggests diagonalizing $P$ by orthogonal rotations as $P=O_PP_dO_P^{-1}, \, \, P_d=diag(p_1,\ldots,p_m)>0$ and  integrating out the associated orthogonal matrices (which incurs the change of measure $dP\to dP_d\prod_{\alpha<\beta}|p_{\alpha}-p_{\beta}|$). We can thus restrict our analysis to finding the stationary point of
\be \label{Lnew}
L(M)|_{R=0} = \Phi_n(Q) + \sum_{\alpha=1}^m  {\cal L}_Q(p_\alpha) + \frac{B''(0)}{8} (\sum_{\alpha=1}^m p_\alpha)^2
\ee
where we have defined
\be\label{FreeEnergy}
\Phi_n(Q):=-\frac{1}{2}\tr \log{Q} + \frac{\beta \mu}{2}\, \tr Q-\frac{\beta^2}{2}\sum_{a,b=1}^n B\left(\frac{Q_{aa}+Q_{bb}-2Q_{ab}}{2}\right)
\ee
 and
\be\label{action2}
 {\cal L}_Q(p):=\quad \frac{i\left(\lambda-\mu_{eff}\right)}{2} p+
\frac{B''(0)}{4}p^2- \frac{1}{2} \ln{p}
\ee
where $\mu_{eff}$ is defined via
\be\label{muef}
\mu_{eff}:=\mu-\beta\sum_{a=1}^nB'\left(\frac{Q_{aa}+Q_{11}-Q_{a1}}{2}\right)
\ee

We can now change the order of integrations over $P_d$ and $Q$ and
consider first the integration over $P_d=diag(p_1,\ldots,p_m)$ for a fixed value of $Q$,
i.e. find the stationary point w.r.t. $P$ at fixed $Q$.
 The saddle point condition in $p_{\alpha}$ yields for all $\alpha=1,\ldots,m$
\be\label{spm}
\frac{\partial}{\partial p_{\alpha}} \left[{\cal L}_Q(p_{\alpha})+\frac{B''(0)}{8} (\sum_{\alpha=1}^m p_\alpha)^2\right]
\ee
\[
=\frac{1}{2}\left[i\left(\lambda-\mu_{eff}\right)+B''(0)\left(p_{\alpha}+\frac{1}{2}\sum_{\alpha}p_{\alpha}\right)-\frac{1}{p_{\alpha}}\right]=0
\]
 We should now remember that (i) we are interested in the limit where $\lambda$ is real (${\rm Im} \, \lambda=0^-$) and
(ii) we have the original constraint $p_{\alpha}>0, \,\, \forall \alpha=1,\ldots,m$. The necessity to deform the integration contour to pass correctly through the stationary points then makes inavoiding to choose all $p_{\alpha}$ equal and such that
 ${\rm Re} \, p_{\alpha}>0$ (see e.g. \cite{F2002a} for a discussion) so that the only contributing saddle-point is
independent of the index $\alpha$:
\be\label{saddle-p}
p_{\alpha}=\frac{ - i\left(\lambda-\mu_{eff}\right)+\sqrt{4B_m''(0)-\left(\lambda-\mu_{eff}\right)^2}}{2B_m''(0)}:=p^{(m)}_{+}, \quad \forall \alpha=1,\ldots,m
\ee
where we denoted $B_m''(0):=B''(0)\left(1+\frac{m}{2}\right)$. Note that in the replica limit $m\to 0$ we have simply $B_m''(0)=B''(0)$.

At the next step we see that in the action Eq.(\ref{Lnew}) taken at the saddle point for $P$, the term
$\sum_{\alpha=1}^m{\cal L}_Q(p_{\alpha})=m{\cal L}_Q\left(p^{(m)}_{+}\right)$ is proportional to $m$,
and the last term in Eq.(\ref{Lnew}) is proportional to $m^2$. If we now
look for the saddle point for $Q$, and we consider $m$ tending to zero before $n$ does,
the only term which couples $P$ and $Q$, $m{\cal L}_Q(p_{+}) + \frac{B''(0)}{8} m^2 \left(p^{(m)}_+\right)^2$, cannot modify the saddle-point value for the matrix $Q$. The latter therefore should be
evaluated by extremizing $\Phi_n(Q)$ only. We thus see that the two extremization procedures essentially decouple, which allows
to perform the two replica limits $m\to 0$ and $n\to 0$ essentially independently.
The same conclusion is reached if one solves the coupled saddle point equations
for $P,Q$ in the limit $m \to 0$, rather than sequentially as above.

From Eq.(\ref{Gl1new3}) and the above saddle point analysis we now conclude that
\be
\lim_{N\to \infty} \overline{\left\langle G(\lambda, {\bf u})\right\rangle_T} = i \, p^{(m=0)}_{+}
\ee
Using Eq.(\ref{rhoT}) this allows to extract the mean spectral density of the Hessian
eigenvalues at a temperature $T$, and we find
\be\label{semi}
 \rho_{T}(\lambda) = 8 \frac{ \sqrt{(\lambda_+-\lambda) (\lambda-\lambda_-)} }{\pi( \lambda_+-\lambda_-)^2} \quad , \quad
\int_{\lambda_-}^{\lambda_+} \rho_{T}(\lambda) d\lambda = 1
\ee
where the lower and upper spectral edges given by
\be\label{semi1}
\lambda_{-} = \mu_{eff}- 2 \sqrt{B''(0)}, \quad \lambda_{+} = \mu_{eff}+ 2 \sqrt{B''(0)}
\ee
and $\mu_{eff}$ defined in Eq.(\ref{muef}). Hence the spectral density of the Hessian at the global minimum, $\lim_{T \to 0}\rho_{T}(\lambda)$ to leading order at large $N$,
{\it is a semi-circle}, with two parameters, its center and its width. To determine those parameters one needs to solve the
{\it equilibrium} optimization problem, by extremizing $\Phi_n(Q)$ at a given inverse temperature $\beta$ and eventually taking $n\to 0$. The resulting values for elements $Q_{ab}$ of the optimal matrix $Q$ in the limit $\beta \to \infty$ should be then substituted to Eqs.(\ref{muef}) and (\ref{semi1}). Fortunately, the replica extremization problem arising here is exactly the same as solved in the various works
described in Section \ref{sec:high} in the framework of the Parisi scheme of
Replica Symmetry Breaking (RSB), and we will refer to mainly \cite{FS2007} and \cite{LDMW}
when discussing this solution in the next section.

 \subsection{Solution of the extremization problem in the Parisi RSB scheme}

 The scheme starts with a standard assumption that in the replica limit $n\to 0$ the integral is dominated by configurations
of matrices $Q$ which for finite integer $n$ have a special hierarchically built
structure characterized by the sequence of integers
\begin{equation}\label{parisiseq}
n=m_0\ge m_1\ge m_2\ge \ldots\ge m_k\ge m_{k+1}=1
\end{equation}
and the values placed in the off-diagonal entries of the $Q$ matrix block-wise, and
satisfying:
\begin{equation}\label{parisiseq1}
0<q_0\le q_1\le q_2\le \ldots\le q_k
\end{equation}
Finally, we complete the procedure by filling in the $n$ diagonal
entries $Q_{aa}$ of the matrix $Q$ with one and the same
value $Q_{aa}=q_d:=q_{k+1}\ge q_k$.
The subsequent treatment is much facilitated by
introducing the following (generalized) function of the variable
$q$:
\begin{equation}\label{xstep}
x(q)=n+\sum_{l=0}^k (m_{l+1}-m_l)\,\theta(q-q_l)
\end{equation}
where we use the notation $\theta(z)$ for the Heaviside step
function: $\theta(z)=1$ for  $z>0$ and zero otherwise. In view of
the inequalities Eq.(\ref{parisiseq},\ref{parisiseq1}) the
function $x(q)$ is piecewise-constant non-increasing, and changes
between $x(0<q<q_0)=m_0\equiv n$ through $x(q_{i-1}<q<q_i)=m_i$ for $i=1, \ldots,k$ to finally $x(q_k<q<q_d)=m_{k+1}\equiv 1$.
In the replica limit $n\to 0$ this function becomes a nontrivial non-increasing function of the variable $q\in[q_0,q_k]$, and outside that interval $x(0<q<q_0)=0$ and $x(q_k<q<q_d)=1$.
One can now express the eigenvalues of any function $g(Q)$ of the hierarchical matrix $Q$ in terms of simple integrals involving $x(q)$ (see e.g. \cite{CS,MP1,MP2})
 and eventually represent the replica limit of eq.(\ref{FreeEnergy}) as a variational functional of $x(q)$
and the parameters $q_d,q_0,q_k$: $\lim_{n\to 0}\frac{1}{n}\Phi_n(Q)=\Phi_{T}[x(q);q_d,q_0,q_k]$.
The explicit form of the functional is given, e.g., in Eq.(47) of \cite{FS2007} (we note that in the paper \cite{FS2007} the covariance $B(q)$ is denoted $f(q)$, and the regularization parameter $\epsilon$ is denoted $a^2$).
Similarly, it is easy to show that one can express the combinations entering Eq.(\ref{muef}) in terms of the same objects as
\be\label{muef1}
\mu_{eff}=\mu- \frac{1}{T} \left[B'(0)-B'(q_d-q_k)-\int_{q_0}^{q_k}B''(q_d-q)x(q)dq \right]
\ee

By varying the resulting functional $\Phi_{T}[x(q);q_d,q_0,q_k]$ one obtains a system of stationarity equations, which may have different types of solutions which we briefly recapitulate below.

\subsubsection{Replica-symmetric solution.}

In the range of parameters in the $(\mu,T)$ plane satisfying the inequality $\mu > \sqrt{B''(T/\mu)}$ the relevant stable solution
to the stationarity equation is of the simplest, {\it replica-symmetric} form such that $q_0=q_k$ so that the function $x(q)$
has  no nontrivial support. The parameters $q_k$ and $q_d$ can be explicitly found:
\be\label{RS}
q_d=\frac{T}{\mu}-\frac{1}{\mu^2}B'\left(\frac{T}{\mu}\right), \quad q_0=-\frac{1}{\mu^2}B'\left(\frac{T}{\mu}\right)
\ee
In terms of them the combination $\mu_{eff}$ from Eq.(\ref{muef1}) is given by
\be\label{muefRS}
\mu_{eff}=\mu- \frac{1}{T} \left[B'(0)-B'(q_d-q_0)\right]
\ee
Combining Eq.(\ref{RS}) and Eq.(\ref{muefRS}) we see that $\mu_{eff}$ has a well-defined zero-temperature limit given by
\be\label{muefRSzero}
\lim_{T \to 0} \mu_{eff} = \mu^{(0)}_{eff}=  \mu+\frac{1}{\mu}B''(0)
\ee
This limit obviously makes sense only as long as it is performed in the phase where the RS solution remains stable at zero temperature, i.e. for $\mu$ satisfying  $\mu>\mu_{c}=\sqrt{B''(0)}$. Substituting this value to Eq.(\ref{semi1}) we find that the lowest edge of the semicircle is given by
\be\label{lowedgeRS}
\lambda^{(RS)}_{-}=\mu+\frac{1}{\mu}B''(0)-2\sqrt{B''(0)}=\frac{(\mu-\mu_c)^2}{\mu}>0, \quad \mu>\mu_{c}=\sqrt{B''(0)}
\ee
We conclude that the lowest eigenvalue of the Hessian at the global minimum in this case is separated by a finite gap.
The width of this gap tends to zero when one approaches the critical value $\mu_c$, known to separate the 'simple' type of landscape (i.e. one with
typically a single minimum and no other saddle points) for $\mu>\mu_c$ from the 'complex' landscapes characterized by both exponentially
many minima and saddle points of all indices \cite{Fyo04,FyoWil07,FyoNad12} for $\mu<\mu_c$. Under these circumstances the replica-symmetric solution loses its stability via the so-called de Almeida-Thouless instability \cite{AT}, and needs to be replaced by a different solution.
The nature of the latter crucially depends on the type of correlations: short-ranged (SRC) vs. long-ranged (LRC).

\subsubsection{1-step replica symmetry breaking solution: SRC potentials.}

In the case of SRC potentials described in Eq.(\ref{shortranged}) the correct solution for $\mu<\mu_c$ corresponds to the simplest scheme of
the replica symmetry breaking scheme, the so-called 1RSB, which is characterized by a step-like shape of the function $x(q)$:
\be\label{1RSB}
x(q)=\left\{\begin{array}{cc} 0, & q\in [0,q_0) \\ m, & q\in[q_0,q_1] \\ 1, & q\in (q_1,q_d]\end{array} \right.
\ee
where $m$ is a temperature-dependent parameter to be found from the variation procedure (note that the parameter $m$ used everywhere in this section should not be confused  with the number of replica $m$ used in the previous sections).
Following \cite{FS2007} we find convenient to re-express the parameters $q_1$ and $q_d$ in terms of combinations
\be\label{1RSBparam}
{\cal Q}=q_1-q_0, \quad y=q_d-q_1,
\ee
 In these terms we find for $\mu_{eff}$ from Eq.(\ref{muef1}) the following representation:
 \be\label{muef1RSB}
\mu_{eff}=\mu-\frac{1}{T}\left[B'(0)-B'(y)\right]+ \frac{m}{T} \left[B'(y+{\cal Q})-B'(y)\right]
\ee
Moreover, the stationarity conditions imply the relation $y=\frac{T}{\mu}-m\,{\cal Q}$ (see Eq.(74) in \cite{FS2007}).
To perform the zero-temperature limit one first realizes that then $m\to 0$ in such a way that the ratio $m/T$ tends to a finite limit, i.e. $\lim_{T\to 0}\frac{m}{T}=v<\infty$, then the parameter ${\cal Q}$ tends to a finite value
at $T=0$.

 Using these facts it is straightforward to perform the $T\to 0$ limit in
 Eq.(\ref{muef1RSB}) and obtain
 \begin{equation}\label{muef1RSBzero}
 \mu^{(0)}_{eff}=\mu+\frac{1}{\mu}B''(0)+v\left[B'({\cal Q})-B'(0)-B''(0)\,{\cal Q}\right]
 \end{equation}
 so that for the lower edge of the Hessian spectrum we have
 \be\label{lowedge1RSB}
\lambda^{\rm (1RSB)}_{-}=\frac{(\mu-\mu_c)^2}{\mu}+v\left( B'({\cal Q})-B'(0)-B''(0)\,{\cal Q}\right), \quad \mu<\mu_{c}=\sqrt{B''(0)}
\ee

Below we show in full generality that
 \be\label{pos}
\lambda^{\rm (1RSB)}_{-} \ge 0
\ee
 Note that it  is not immediately
obvious to see that Eq.(\ref{lowedge1RSB}) satisfies Eq.(\ref{pos}) due to  $v>0$ and $\Delta:=B'({\cal Q})-B'(0)-B''(0) {\cal Q} <0 \, , \, \forall {\cal Q}>0$ for SRC covariances. As an example one can consider
$B({\cal Q})=e^{-{\cal Q}}$ for which $\Delta=1-{\cal Q} -e^{-{\cal Q}}<0$ for ${\cal Q}>0$. To ascertain the sign of $\lambda^{(1RSB)}_{-}$ one needs to study the system of two closed equations satisfied
 by $v$ and ${\cal Q}$ everywhere for $\mu<\mu_c$ ( Eqs.(89)-(90) in Eq.\cite{FS2007}), presented in the Introduction,
 see Eq.(\ref{RSBQv1}) and Eq.(\ref{RSBQv2}).

Now one can
use Eq.(\ref{RSBQv1}) to eliminate $B'({\cal Q})-B'(0)$ in Eq.(\ref{lowedge1RSB}).
After some rearranging one finds
\be\label{lowedge1RSBnew}
  \lambda^{\rm (1RSB)}_{-}= \frac{\mu}{1- \mu v {\cal Q}} + \frac{1-\mu v {\cal Q}}{\mu}\, B''(0)  -2\sqrt{B''(0)}
\ee
\be\label{lowedge1RSBnew1}
=\left(\sqrt{\frac{\mu}{1- \mu v {\cal Q}}}-\sqrt{ \frac{1-\mu v {\cal Q}}{\mu}\, B''(0)}\right)^2 \
=\frac{\mu_c^2}{\mu(1-\mu v {\cal Q})}\,D^2\ge 0
\ee
where we denoted
\be\label{lowedge1RSBnew2}
D=\frac{\mu}{\mu_c}-1+\mu\,v\,{\cal Q}
\ee

The above expression can be used to study the critical behaviour of the spectral gap $ \lambda^{\rm (1RSB)}_{-}$ just below the threshold $\mu=\mu_c$. Note that the system of equations (\ref{RSBQv1})-(\ref{RSBQv2}) is obviously always satisfied by ${\cal Q}=0$ corresponding to the replica-symmetric solution,
  but we are interested in finding
 a nontrivial solution with ${\cal Q}>0$.  For a general $\mu<\mu_c$, and general $B(q)$, this can be done only numerically: see  Fig. 1 in the Introduction  for a plot of the lower edge of the Hessian for one example of a SRC potential.
 Close to the critical value however, when $\delta=1-\frac{\mu}{\mu_c}\ll 1$
 one can develop the perturbation theory in $0<\delta\ll 1$. Since for $\delta\to 0$ we must have ${\cal Q}\to 0$  we look for a solution in the form:
 \begin{equation}\label{perturb}
v=v_0+\delta v_1+\delta^2v_2+\ldots, \quad {\cal Q}=Q_1\delta+Q_2\delta^2+Q_3\delta^3+\ldots , \quad \mbox{for}\,\, \mu=\mu_c(1-\delta)
 \end{equation}
 where $\mu_c=\sqrt{B''(0)}$.  To find these expansion coefficients we substitute Eq.(\ref{perturb}) into
   Eqs.(\ref{RSBQv1})-(\ref{RSBQv2}) and get for the two leading orders:
  \be\label{firstordersol}
   v_0=-\frac{B'''(0)}{2\mu_c^{3}}, \quad Q_1=-\frac{2\mu_c^2}{B'''(0)}
\ee
\be \label{secondordersol}
\fl  v_1=-\frac{\frac{3}{2}[B'''(0)]^2-B''(0)B^{''''}(0)}{2\mu_c^{3}B'''(0)},
 \quad Q_2=B''(0)\frac{3[B'''(0)]^2-10B''(0)B^{''''}(0)}{6\left[B'''(0)\right]^{3}}
 \ee
Substituting the expansion Eq.(\ref{perturb}) into Eq.(\ref{lowedge1RSBnew2}) we see that
\be\label{D1}
D = -\delta\left[1-\mu_cv_0Q_1\left(1+\delta\left\{\frac{v_1}{v_0}+\frac{Q_2-Q_1}{Q_1}\right\}\right)\right]+O(\delta^3)
\ee
Now, Eq.(\ref{firstordersol}) implies $\mu_cv_0Q_1=1$ and further using Eq.(\ref{secondordersol})
we arrive at
\be\label{D2}
D = \frac{1}{6}\,\frac{\frac{3}{2}[B'''(0)]^2-B''(0)B^{''''}(0)}{[B'''(0)]^2}\, \delta^2 + O(\delta^3)
\ee

In view of Eq.(\ref{lowedge1RSBnew1}) we then conclude that the first nonvanishing contribution
to $ \lambda^{(1RSB)}_{-}$ is order $O(\delta^4)$, and is explicitly given by Eq.(\ref{lowedge1RSBperturb1}) presented in the Introduction.

Moreover, using the discussion after Eq.(\ref{Schwarziancriterion}) we see that the leading term in Eq.(\ref{lowedge1RSBperturb1}) is {\it strictly positive} in the case of all short-range potentials, and vanishes only for the logarithmically-correlated case Eq.(\ref{logcorr}). We will show in the end of our consideration, see Eq. (\ref{lowedgeFRSB}) below, that such vanishing is not a coincidence: for the logarithmic case the spectral gap $ \lambda^{(1RSB)}_{-}$ is {\it identically zero} everywhere in the phase with broken replica symmetry, $\mu<\mu_c$, so that the associated spectrum is {\it gapless}. In contrast, for a generic SRC potential the spectrum in the 1RSB phase is gapped, and tends to zero as proportional to $(\mu_c-\mu)^4$ on approaching the transition point $\mu_c=\sqrt{B''(0)}$, which is much faster than from the RS side, when vanishing is proportional to $(\mu-\mu_c)^2$, see Fig. 1.

 \subsubsection{Full replica symmetry breaking solution: LRC potentials.}
 LRC potentials characterized by Eq.(\ref{longranged})\footnote{
Note that in the paper \cite{FS2007} the long-ranged potentials were formally considered without the short-scale regularization, i.e. for $\epsilon=0$. This resulted in $B''(0)=\infty$ and hence the replica-symmetry was fully broken at zero temperature at any value of $\mu$.
For the purposes of the present study keeping a finite regularization is essential.} define the most complex type of the Gaussian landscapes and are desribed
 by a pattern of full replica symmetry breaking (FRSB), with a nontrivial continuous non-increasing function $x(q)$ in
 the whole nonvanishing interval $q\in[q_0,q_k]$. This function can be found explicitly (see Eq.(53) of \cite{FS2007}):
 \be\label{FRSB}
 x(q)=\left\{\begin{array}{cc} 0, & q\in [0,q_0) \\ -\frac{T}{2}\frac{B'''(q_d-q)}{\left[B''(q_d-q)\right]^{3/2}}, & q\in[q_0,q_k] \\ 1, & q\in (q_k,q_d]\end{array} \right.
 \ee
 which allows to represent the parameter $\mu_{eff}$ in Eq.(\ref{muef1}) in the form
 \begin{equation}\label{muefFRSB}
 \mu_{eff}=\mu-\sqrt{B''(q_d-q_0)}+\sqrt{B''(q_d-q_k)}-\frac{1}{T}\left[B'(0)-B'(q_d-q_k)\right]
 \end{equation}
 It can be further shown that the parameters $q_0,q_k$ and $q_d$ in FRSB phase satisfy the following relations (see Eqs.(54) $\&$ (56) in \cite{FS2007})
 \be\label{equilcond}
 q_d-q_k=\frac{T}{\sqrt{B''(q_d-q_k)}}, \quad \mu^2=B''(q_d-q_0)
 \ee
 In the low-temperature limit $q_d-q_k\approx T/\sqrt{B''(0)}$ implying that
 \[
 \lim_{T\to 0}\frac{B'(0)-B'(q_d-q_k)}{T}=-\sqrt{B''(0)}
 \]
 which in turn results in the zero-temperature expression
 \begin{equation}\label{muefFRSBzero}
 \mu^{(0)}_{eff}=\mu-\sqrt{B''(q_d-q_0)}+2\sqrt{B''(0)}=2\sqrt{B''(0)}
 \end{equation}
 where we used second of Eq.(\ref{equilcond}). We immediately see from Eq. (\ref{semi1}) that in such a situation the spectrum
 of the Hessian is {\it gapless} for any $\mu<\mu_c$:
 \be\label{lowedgeFRSB}
\lambda^{(FRSB)}_{-}=\mu^{(0)}_{eff}- 2 \sqrt{B''(0)}=0, \quad \mu<\mu_{c}=\sqrt{B''(0)}
\ee
 Obviously, such conclusion corroborates with the standard picture of huge degeneracy in the ground state of FRSB phase,
  indicating presence of 'marginally stable' spatial directions where one can move away from the global minimum almost at no
  increase of the landscape height.

  \subsubsection{ Logarithmically-correlated potential.}

  Finally, let us consider the boundary case of logarithmically-correlated potential, Eq.(\ref{logcorr})
   which although described by 1-step SRB, is expected to share many features with
  FRSB systems ( by this reason this case is sometimes called the "marginal 1RSB"). Going back to 1RSB formulas, we note that the parameters $v$ and $Q$ characterizing the zero-temperature solution can be found explicitly (see Eq.(64) in \cite{FS2007}), for $\mu<\mu_c$
  \be\label{logsol}
  v=\frac{1}{g}, \quad Q=\frac{g}{\mu}-\epsilon
  \ee
   whereas $B'(q)=-g^2/(q+\epsilon)$ so that $B'(0)=-g^2/\epsilon$ and $B''(0)=g^2/\epsilon^2$. Substituting these values to
   Eq.(\ref{lowedge1RSB}) we see that in such a case
  \be\label{lowedgeFRSB}
\lambda^{\rm (log)}_{-}=\left\{\begin{array}{cc}\frac{(\mu-\mu_c)^2}{\mu}>0, & \mu>\mu_{c}=\frac{g}{\epsilon} \\ 0,  & \mu<\mu_{c}=\frac{g}{\epsilon} \end{array}\right.
\ee
 exactly like in FRSB case. Note that the critical value of the confinement $\mu_c$ crucially depends on the short-ranged
regularization $\epsilon>0$.

\section{Conclusions and Discussions}

 In this paper we adapted the replica trick to derive the mean eigenvalue density for the Hessian associated with the global minimum of one of the most paradigmatic models of high-dimensional landscapes: an isotropic homogeneous Gaussian-distributed random potential superimposed on a simple parabolic background with the curvature parameter $\mu>0$, see Eq.(\ref{landscape}).
We  showed that the Hessian eigenvalue density is always of the semicircular form. The position of the lowest edge of the semicircle was found to reflect  the nature of the underlying potential controlled by both the value of the curvature $\mu$ of the deterministic part
and of the covariance structure of the random part. Simple landscapes with generically a single minimum are typical for $\mu>\mu_{c}= \sqrt{B''(0)}$, and we show that  for these landscapes the Hessian at the global minimum is always {\it gapped}, with the low spectral edge being strictly positive. The spectral gap vanishes  as $(\mu-\mu_c)^2$ when approaching from above the transition point  $\mu= \mu_{c}$ separating simple  landscapes from ''glassy'' ones, with exponentially abundant minima. For $\mu<\mu_c$
the Hessian spectrum is qualitatively different for 'genuinely complex' and 'moderately complex' landscapes. In the '' genuinely complex'' landscapes with long-ranged power-law correlations the replica symmetry is completely broken and
  the  Hessian  remains gapless for all values of $\mu<\mu_c$, indicating presence of 'marginally stable' spatial
 directions, as was anticipated in other models with fully broken replica symmetry, see \cite{MPV,MLC2006} and a discussion in the introduction. In contrast, the 'moderately complex' landscapes are typical for short-range correlated random potentials and correspond to the 1-step replica-symmetry breaking mechanism. Their Hessian spectra turn out to be in general gapped, with the gap vanishing on approaching $\mu_c$ from below  with a larger critical exponent, as $(\mu_c-\mu)^4$, see Eq.(\ref{lowedge1RSBperturb1}) and the Figure 1. To the best of our knowledge this type of critical behaviour has not yet been discussed previously. It would be therefore interesting to see if this feature is generic for other 'topological trivialization' transitions associated with the 1RSB mechanism (e.g. one induced by increasing magnetic field in the spherical model, see e.g. \cite{FyoLeD14,Fyo16,RBBC2018} and references therein), and what are the  associated changes of the landscape shape around the global minimum which is responsible for such an asymmetry of the lowest Hessian eigenvalue. We leave this and related problems for a future investigation.

Finally, we demonstrated that the potentials with {\it logarithmic} correlations,  which are at the
frontier between 1RSB and FRSB (they are called marginal 1RSB), share both the 1RSB nature and a gapless spectrum.

 The method used here is quite versatile and opens the way to explore further questions.
In particular, it should be possible to study 'bulk' spectral correlations (presumably, of Wigner-Dyson type), and further fluctuations of (delocalized) eigenvectors of the Hessian  within a variant of our method involving 'superbosonisation' approach to resolvents \cite{LSZ,Sommers}. What concerns the few lowest eigenvalues of
the Hessian, those would be natural to expect to show Tracy-Widom type fluctuations in the gapped phases, whereas at the transition point and in the gapless phase their statistics should be quite different. This will require studying finite $N$ corrections, and
one may wonder what remains of the present picture at finite $N$, especially in the interesting
case of logarithmic potentials where the RSB transition is known to survive (and for which, in fact, infinite and finite $N$ share a lot of similarities) \cite{CLD,FB2008a,logmultlec}. Obtaining a more detailed
picture of the low lying states and their associated Hessian would also be of great
interest. Finally, extensions to pinned manifolds are presently under investigation.

\bigskip

{\bf Acknowledgments:}  This research was initiated during YVF visit to Paris supported by
the Philippe Meyer Institute for Theoretical Physics, which is gratefully acknowledged.
The research at King's College (YVF) was supported by  EPSRC grant  EP/N009436/1 "The many faces of random characteristic polynomials".
PLD acknowledges support from ANR grant ANR-17-CE30-0027-01 RaMa-TraF.

\appendix

\section{Appendix: $O(N)-$invariant integration formula.}
\label{app1}

In this Appendix we recall some identities which are very helpful when
dealing with integrals with $O(N)$-invariant integrands. A statement equivalent to one of the Theorem I below seemingly first appeared in \cite{Percus87}, and was later on rediscovered in \cite{F2002a}. A detailed proof is provided in
Appendix B of \cite{logmultlec} (see also Appendix A of
\cite{DavidInvariantIntegral}).\\

{\bf Theorem I}

{\sf
Consider a function $ F({{\bf u}_1},...,{\bf{u}_n})$
of $N$-component real vectors ${\bf u}_a$ with $ 1\le a\le n$ such that
\begin{equation}\label{conv1}
\int_{R^N}d{\bf u}_1...\int_{R^N}d^2{\bf u}_n
|F({\bf u_1},...,{\bf u_n})|<\infty
\end{equation}
Denoting by ${\bf u}^T$ the vector transposition
suppose further that the function $F$ is invariant with resect to any global $O(N)$ rotation: ${\bf u}_a\to O{\bf u}_a, \forall a=1,\ldots n$. Then necessarily it depends only on $n(n+1)/2$ scalar products ${\bf u}^{T}_{a}
{\bf u}_{b}\,\, 1\le a,b\le n$ and can be therefore rewritten as
a function $F({\bf u_1},...,{\bf u_n}) = {\cal F}(\tilde{Q})$ of the
$n\times n$ real symmetric matrix $\tilde{Q}$:
\[
\tilde{Q}_{a,b}
={\bf u}^{T}_{a}{\bf u}_{b}\left.\right|_{1\le a,b\le n}
\]
Then for $N>n$ the integral defined as
\begin{equation}\label{defint}
I_{N,n}=\int_{R^N}d{\bf u}_1... \int_{R^N}d{\bf u}_n
F({\bf u}_1,...,{\bf u}_n)
\end{equation}
is equal to
\begin{equation}\label{trans1}
I_{N,n}={\cal C}^{(o)}_{N,n}
\int_{Q\ge 0} dQ \, \left( \det{Q} \right)^{(N-n-1)/2}
{\cal F}(Q)
\end{equation}
where the proportionality constant is given by
\begin{equation}\label{propconst}
{\cal C}^{(o)}_{N,n}= \frac{\pi^{\frac{n}{2}\left(N-\frac{n-1}{2}\right)}}
{\prod_{k=0}^{n-1}\Gamma\left(\frac{N-k}{2}\right)}
\end{equation}
and the integration in Eq.(\ref{trans1})
goes over the manifold
of real symmetric positive semidefinite $n\times n$ matrices $Q\ge 0$,
with measure $dQ = \prod_{a \leq b} dQ_{ab}$.
}\\

One of the ways of proving the Theorem is to exploit the following matrix integral identity:
\begin{equation}\label{J1}
 {\cal J}_{n,N,\epsilon}(Q)=\int e^{-\frac{i}{2}tr[QX]} \frac{1}{\det\left(\epsilon {\bf 1}_n-iX\right)^{N/2}}\,dX,
 \end{equation}
\begin{equation}\label{J2}
=\frac{1}{2^{\frac{Nn}{2}}\pi^{\frac{n(n-1)}{4}}}
\frac{1}{\prod_{j=0}^{n-1}\Gamma\left(\frac{N-j}{2}\right)}
\, \det{Q}^{\frac{N-n-1}{2}}\,
e^{-\frac{1}{2}\epsilon\tr\,Q}\,\, \prod_{j=1}^n\theta(q_j)
\end{equation}
where the integration goes over $n\times n$ real symmetric matrices: $X_{ab}=X_{ba}$ with the volume element $dX=\prod_{a}\frac{dX_{aa}}{4\pi}\prod_{a<b}\frac{dX_{ab}}{2\pi}$, $\epsilon>0, N\ge n+1$, $\Gamma(\nu)$ is the Euler Gamma-function, $Q$ is $n\times n$ real symmetric and $\theta(x)=1$ for $x\ge 0$ and zero otherwise.  The identity
can be proved \cite{F2002a,logmultlec} following a method suggested for evaluating a related integral due to Ingham and Siegel \cite{Bellman}.

Finally, let us mention that there exists a useful generalization of the Theorem I, known as the 'superbosonisation', which extends it to the case when vectors ${\bf u}$ are replaced with {\it supervectors} involving both commuting ('bosonic') and anticommuting ('fermionic') entries, see an informal account in \cite{Sommers}.

\section{Weak confinement limit $\mu\to 0$ for SRC potentials}
\label{app2}

Here we consider the case of SRC potentials, with $q B'(q) \to 0$ as $q \to +\infty$ which is generic but excludes the marginal case of logarithmically-correlated potential. Under this condition
 we establish the leading small $\mu$ behavior of the lowest Hessian edge.

In the limit $\mu \to 0$ the equations \eqref{RSBQv1}-\eqref{RSBQv2}
lead to ${\cal Q}, v \to +\infty$, $1- \mu\,v\,{\cal Q}\ \to 0$ with, keeping only the leading terms
\be
 \frac{\mu^2\,{\cal Q}}{1-\mu\,v\,{\cal Q}} \simeq -B'(0) \quad , \quad
 \ln{\left(1-\mu\,v\,{\cal Q}\right)} \simeq -1 - B(0) v^2
\ee
Multiplying the first equation by $v/\mu$, and denoting $X=1-\mu v Q$
we obtain
\be
\frac{1}{X} \simeq 1- v \frac{B'(0)}{\mu} \quad , \quad \ln \frac{1}{X} = 1+ v^2 B(0)
\ee
Introducing $y:=v\sqrt{B(0)}\gg 1$ we then have from the above system:
\be
y^2\approx \ln{\frac{1}{e \mu}}+\ln{y}+\ln{\frac{-B'(0)}{\sqrt{B(0)}}}+O(1/y)
\ee
which is solved asymptotically for $\mu\to 0$ by
\be
y \approx \sqrt{\ln{\frac{1}{e \mu}}}+\frac{1}{4}\frac{\ln{\ln{\frac{1}{e \mu}}}}{\sqrt{\ln{\frac{1}{e \mu}}}}
\ee
which finally gives for the gap from Eq.(\ref{lowedge1RSBnew})
\bea
&& \lambda^{\rm (1RSB)}_{-} = \frac{\mu}{X} + \frac{X}{\mu}B''(0) - 2 \sqrt{B''(0)} \simeq -\frac{B'(0)}{\sqrt{B(0)}}\,y - 2 \sqrt{B''(0)} \\
&& ~~~~~~~~\simeq \frac{-B'(0)}{\sqrt{B(0)}} \left[\sqrt{\ln \frac{1}{e \mu}}+\frac{1}{4}\frac{\ln{\ln{\frac{1}{e \mu}}}}{\sqrt{\ln{\frac{1}{e \mu}}}}\right] - 2 \sqrt{B''(0)}
\eea

\end{document}